\documentclass[conference]{IEEEtran}
\usepackage{cite}
\usepackage{amsmath,amssymb,amsfonts}
\usepackage{algorithmic}
\usepackage{graphicx}
\usepackage{textcomp}
\usepackage{xcolor}
\usepackage{tabularx}
\usepackage{booktabs}
\usepackage{makecell}
\usepackage{rotating}
\usepackage{tabularray}
\usepackage{multirow}
\usepackage{vcell}
\usepackage{diagbox}
\usepackage{array}
\def\BibTeX{{\rm B\kern-.05em{\sc i\kern-.025em b}\kern-.08em
    T\kern-.1667em\lower.7ex\hbox{E}\kern-.125emX}}

\usepackage{tikz}
\usepackage{textcomp}
\usepackage{hyperref}

\newcommand\copyrighttext{%
  \footnotesize \textcopyright 2023 IEEE. Personal use of this material is permitted.
  Permission from IEEE must be obtained for all other uses, in any current or future
  media, including reprinting/republishing this material for advertising or promotional
  purposes, creating new collective works, for resale or redistribution to servers or
  lists, or reuse of any copyrighted component of this work in other works.
  DOI: \href{https://doi.org/10.1109/DAPPS57946.2023.00017}{10.1109/DAPPS57946.2023.00017}}
\newcommand\copyrightnotice{%
\begin{tikzpicture}[remember picture,overlay]
\node[anchor=south,yshift=10pt] at (current page.south) {\fbox{\parbox{\dimexpr\textwidth-\fboxsep-\fboxrule\relax}{\copyrighttext}}};
\end{tikzpicture}%
}

\begin{document}

\title{A Taxonomy of Decentralized Identifier Methods for Practitioners}

\makeatletter
    \newcommand{\linebreakand}{%
    \end{@IEEEauthorhalign}
    \hfill\mbox{}\par
    \mbox{}\hfill\begin{@IEEEauthorhalign}
}

\newcommand{\thickhline}{%
    \noalign {\ifnum 0=`}\fi \hrule height 1.4pt
    \futurelet \reserved@a \@xhline
}
\newcolumntype{"}{!{\vrule width 1pt}}

\newcolumntype{[}{@{\vrule width 1pt\hspace{6pt}}}
\newcolumntype{]}{@{\hspace{6pt}\vrule width 1pt}}

\makeatother

\author{
\IEEEauthorblockN{Felix Hoops}
\IEEEauthorblockA{\textit{Department of Computer Science} \\
\textit{Technical University of Munich}\\
Munich, Germany \\
felix.hoops@tum.de}
\and
\IEEEauthorblockN{Alexander Mühle}
\IEEEauthorblockA{\textit{Hasso-Plattner-Institute} \\
\textit{University Potsdam}\\
Potsdam, Germany \\
alexander.muehle@hpi.de}
\linebreakand
\IEEEauthorblockN{Florian Matthes}
\IEEEauthorblockA{\textit{Department of Computer Science} \\
\textit{Technical University of Munich}\\
Munich, Germany \\
matthes@tum.de}
\and
\IEEEauthorblockN{Christoph Meinel}
\IEEEauthorblockA{\textit{Hasso-Plattner-Institute} \\
\textit{University Potsdam}\\
Potsdam, Germany \\
christoph.meinel@hpi.de}
}

\maketitle

\copyrightnotice

\begin{abstract}
 A core part of the new identity management paradigm of Self-Sovereign Identity (SSI) is the W3C Decentralized Identifiers (DIDs) standard. 
 The diversity of interoperable implementations encouraged by the paradigm is key for a less centralized future, and it is made possible by the concept of DIDs.
 However, this leads to a kind of dilemma of choices, where practitioners are faced with the difficult decision of which methods to choose and support in their applications. 
 Due to the decentralized development of DID method specifications and the overwhelming number of different choices, it is hard to get an overview.
 In this paper, we propose a taxonomy of DID methods with the goal to empower practitioners to make informed decisions when selecting DID methods. To that end, our taxonomy is designed to provide an overview of the current landscape while providing adoption-relevant characteristics.
 For this purpose, we rely on the Nickerson et al. methodology for taxonomy creation, utilizing both conceptual-to-empirical and empirical-to-conceptual approaches.
 During the iterative process, we collect and survey an extensive and potentially exhaustive list of around 160 DID methods from various sources.
 The taxonomy we arrive at uses a total of 7 dimensions and 22 characteristics to span the contemporary design space of DID methods from the perspective of a practitioner.
 In addition to elaborating on these characteristics, we also discuss how a practitioner can use the taxonomy to select suitable DID methods for a specific use case.
\end{abstract}

\begin{IEEEkeywords}
Identity Management, Self-Sovereign Identity, Decentralized Identifiers, DID Methods, Taxonomy
\end{IEEEkeywords}

\section{Introduction}
\label{sec:intro}
With the rising public sensitivity for data hoarders on today’s internet, it is no surprise that we strive to once again evolve our handling of online identity. Early on, every internet service held their own identity data in the form of traditional username and password accounts. But since then, we have mostly moved on to federated accounts. Tech giants such as Google, Facebook, and Apple keep our identity information and negotiate the login with third parties when we request it.

While this is often more convenient, leading to fewer accounts to keep track of for the user, it also dangerously centralizes our personal information. This centralization leads to an array of problems. Naturally, the few big identity providers make for an attractive target for attackers due to the vast amounts of critical data. But even more problematic is the power we give to these identity providers. This starts with the detailed data they can collect about us due to them being our intermediary in so many login procedures from online stores to different social media. That of course means they know all our personal information, such as gender, legal name, date of birth and home address. Furthermore, they know with what services we interact when and from where. All this data is extremely valuable to advertisers, but also to the identity providers themselves, as it usually helps them optimize their service portfolio and thus gives a level of user insight that no other competitor could ever feasibly reach.

Finally, today’s level of centralization of identity providers exposes everyone to the risk of service outages. Either by accident or by design. Any big identity provider having an outage would leave millions of people locked out of their accounts with dozens of third party service providers. And if any identity provider wanted to, or was forced to by a government, to suspend service to certain people or groups of people, they could do so in an instant, disrupting the lives and livelihoods of people and businesses alike.

Decentralizing our global identity management should thus be a central goal for the near future. Two recent W3C standards are taking the spotlight as essential tools to reach this goal: Decentralized Identifiers (DIDs)\footnote{https://www.w3.org/TR/did-core/} and Verifiable Credentials (VC)\footnote{https://www.w3.org/TR/vc-data-model-2.0/}. The former being an identifier standard, while the latter standardizes a format for verifiable claims. Anyone can create censorship resistant, globally unique, universal identifiers in the form of DIDs. Gathering assertions about themselves, bound to a specific DID, allows users to turn an identifier into an identity. Companies can state that someone works for them, universities can confirm someone’s status as a student, and countries can issue fully digital passports. 

DIDs promise a future internet turned upside down: from big identity providers being an intermediary for our every action towards each user being their own identity keeper. Independent of implementation technology, this design concept is known as Self-Sovereign Identity (SSI). Through their independence of third-party actors and their interoperability, DIDs begin to prove themselves as a crucial technology enabling this future where every user can bring their own identity\textemdash or account\textemdash to every digital service they choose to interact with.

This level of interoperability is a key point of this W3C specification and ensures that DIDs can be tailored to specific use cases and that the specification is extendible to take advantage of future technologies and infrastructures. However, currently it might be materializing as a major hurdle to overcome on the way to broad adoption, as numerous DID implementations, called DID methods, are created without much coordination and with vastly different implementation approaches. This massive and diverse number of very different DID methods makes getting an overview very time-consuming. Just compiling a list of them is already a challenge because the few existing lists online are limited in scope and riddled with errors or inconsistencies.

For DIDs to be used effectively, it is vital that system architects can quickly gain an overview of existing DID methods and understand their characteristics and capabilities. They need to be able to make an informed decision on what DID methods to support for a given system and what DID methods to recommend to different stakeholders. To that end, we present a taxonomy of DID methods tailored to assist practitioners in selecting the most suitable one for each use case.

This paper is organized as follows. First, we provide background on the DID standard in Section~\ref{sec:background}. Next, we review related work already done by others in Section~\ref{sec:related} before we go on to describe our methodology to develop our taxonomy in Section~\ref{sec:methodology}. Then, we present our taxonomy for DID selection along with some general insights we have gained developing it in Section~\ref{sec:taxonomy}. We finish by summarizing our results and providing some suggestions for future work in Section~\ref{sec:conclusion}.

\section{Background}
\label{sec:background}
Self-sovereign identity (SSI) is a concept coined by Christopher Allen~\cite{allen2016path} that can be understood as a form of user-centered design in identity management. Users govern their own identity data that they get issued by different issuers. And users decide when they present what data to which verifier. Verifiers would ideally never have to directly contact issuers during the verification procedure. 

Decentralized Identifiers (DIDs) are a key technology to implement SSI-conforming systems. They are carefully designed as a “meta-specification”. That means that the DID standard does not govern how identifiers are managed, but instead how sub-standards to manage identifiers are to be defined. These sub-standards are known as DID methods, and they are developed completely decentralized. Each method defines where data for the identifier is stored and how it is created, read, updated, and deleted. Though, only the ability to create and read is mandatory.

Each DID is a globally unique, cryptographically verifiable identifier. To achieve this, asymmetric encryption is leveraged. Anyone can create a key pair consisting of private and public key. This process works offline. The resulting public key is globally unique and can be shared without compromising the associated private key, given that there are no sufficiently powerful quantum computers in existence. Apart from encryption, asymmetric encryption also enables digital signatures. Encrypting a message with one's own private key produces a secure digital signature. For efficiency reasons, almost every signature scheme encrypts a hash of the message, instead of the message. Distributing that original message, the signature, and the public key allows anyone else to confirm that signature by decrypting the signature with the public key and comparing it to the message hash. While this is an important technical foundation, it is not sufficient to create usable decentralized identity management. DIDs build on this foundation and enable metadata and key management.

A really simple DID that directly uses a public key could look like this:

\begin{center} \emph{did:key:z6MkhaXgBZD...38x74tKLGpbnnEGta2doK} \end{center}

Any DID generally consists of three parts separated by colons: the prefix “did”, a method identifier like “key”, and a method specific part, which is just a public key in this example. This last part can also be extended into multiple method specific parts. For example, that is used to indicate the used infrastructure for DID methods that support multiple ones. This DID based on Ethereum indicates it is registered on a testnet, which is referred to by its ID “0x5”:

\begin{center} \emph{did:ethr:0x5:0xabcab...c39nEG465ta2abe984193675} \end{center}

Essentially, DID methods provide an abstraction layer for an asymmetric key pair. Instead of directly using a key pair and distributing the public key as the identifier, a DID string contains sufficient information to read or “resolve” the identifier into a DID document. Construction of the document may be possible from just the identifier (ref. example 1) or require querying some underlying storage infrastructure, such as the Ethereum blockchain (ref. example 2). This DID document is a JSON document containing one or more public keys associated with the holder, key usage policies, any relevant service endpoints, and potentially more. A big advantage of this design is that it allows key rotations and additional metadata attached to an identifier.

Summarizing, DIDs introduce ground rules for identifiers that are abstracted from asymmetric public keys and then support theoretically infinitely many ways to handle this abstraction through DID methods. Key to understanding the need for structure in the DID space is that the DID specification leaves a lot of room in its ground rules to design valid DID methods. Relying on centralized infrastructure or services (e.g., \emph{did:dns}), fully decentralized infrastructure in the form of a public blockchain (e.g., \emph{did:ethr}), or even no external storage at all (e.g., \emph{did:key}) is possible when designing a DID method. Also, the differences in the set of supported operations and cost can be staggering.

\section{Related Work}
\label{sec:related}
The field of Self-Sovereign Identity (SSI) and especially the standard Decentralized Identifiers (DIDs) are still young. At the time of writing, the W3C VC data model has only been accepted as a W3C recommendation for 3 years, while the DID has only been moved to the recommendations status last year. Understandably, there has been some work structuring and evaluating DID methods, but not at the scale and longevity of results required. Past scientific contributions can be divided into just two categories: first, introduction to and classification of SSI components including DIDs, and second, work on DID evaluation and evaluation criteria. And finally, we acknowledge the work done outside the scientific community by looking at different online sources compiling lists of DID methods.

This first category of works naturally also addresses DIDs and provide general insights into the DID design space, its state, and some concrete challenges arising from its independently developed DID methods.
An early paper by Mühle et al.~\cite{muhle2018survey} examines the state of SSI-compliant identifiers as part of their overview of SSI components. The authors describe DIDs as a high-level naming scheme and draw comparisons to Uniform Resource Names (URNs). They go on to discuss the problems of DID backup and recovery, before briefly discussing different approaches for sharing of public key material and metadata.

In the face of this emerging field and the subsequent academic tackling of the topic, Cucko et al.~\cite{vcuvcko2021decentralized} created an overview of the current research area. For this purpose they applied a systematic mapping methodology to get insights into the makeup of the research area, the trends, associated challenges and subsequent opportunities for future research in the area of Self-Sovereign Identity.

Brunner et al.~\cite{brunner2020did} take a structured look at the W3C DID and Verifiable Credential (VC) standards. As part of that, they discuss several aspects of DID methods. Revocation of a DID in case of theft or loss of key material is brought up. They also mention that tamper-proof timestamping to support expiry dates is difficult to achieve, as the secure timestamping mechanism would have to be reimplemented for every new DID method supporting expiration. They also raise usability concerns, including the issue of key recovery and lack of human-friendly key material.

Lesavre et al.~\cite{lesavre2019taxonomic} discuss blockchain-based identity management following a taxonomic approach. While they mention DIDs, they focus on structuring blockchain identifiers in general, which are only a subset of all DID methods. As part of their work, they discuss the often arising need to register identifiers publicly and provide an overview of different approaches. These range from on-chain registries, over lightweight identifiers without on-chain transactions, to approaches employing unspent transaction outputs (UTXOs) on Bitcoin.

The concept of DIDs has also been investigated critically, such as in the analysis of immunity passport concepts by Halpin~\cite{halpin2020vision}.
Here, the standard of DIDs was evaluated with a focus on security and privacy questions. However, the evaluation was done in a systematic way but rather an unstructured discussion based on the case study of an immunity passport.

Recognizing the challenges that the ever expanding set of DID methods poses, the W3C has started an internal, but publicly accessible, DID rubric\footnote{https://w3c.github.io/did-rubric/}. It presents a large set of criteria that can be applied to a DID method to evaluate and characterize it. The rubric however is still arguably a difficult starting point, even for a software architect. It assumes that a practitioner already has an overview of the DID method space and is capable of pre-selecting several methods to further evaluate with a specific use case in mind. Only a few criteria are evaluated for a very small set of methods to provide some examples.

Similarly, during the 2022 Rebooting Web of Trust workshop, an evaluation methodology for DID methods was proposed by Cunningham et al.~\cite{cunningham2022ready}. This rubric was aimed at product owners, implementers and standards bodies to help them navigate the DID method landscape and gauge maturity of different methods. 

Fdhila et al.~\cite{fdhila2021methods} have used the W3C DID rubric as a basis to take a deeper look at six select DID methods. They acknowledge the wide range of different methods and state they have purposefully chosen very different examples to evaluate in detail. Their work concludes that all the evaluated methods possess their own strengths and weaknesses, justifying their existence.

Finally, it shall be noted that several lists of DID methods are maintained online. All of them include some dead or incomplete methods. The W3C DID Specification Registries\footnote{https://w3c.github.io/did-spec-registries/} compile all known parameters and values used across different methods. However, the document is still a work in progress and advises against using it. Part of this document is a list of DID methods under development that are able to meet the minimum requirements outlined in the DID specification\footnote{https://www.w3.org/TR/did-core/\#methods}. Some minimal structure is provided by also listing what registry infrastructure is used, but the information is incomplete and not always fully accurate.

The Universal Resolver project\footnote{https://github.com/decentralized-identity/universal-resolver} is intended to provide one software that includes resolving capability for as many DID methods as possible. It is open source, and part of the public GitHub repository is a list of supported methods. Next to the method identifier itself, links to the specification, implementation, and some comments are included for most methods.

There are a two more lists online which appear to be compiled by interested third-parties. The Decentralized Identity Web Directory is a website bundling a lot of identity-related resources, including a list of DID methods with links to the specifications\footnote{https://decentralized-id.com/web-standards/w3c/wg/did/decentralized-identifier/}. Then, there is the DID Directory\footnote{https://diddirectory.com/}. It has taken the information from the W3C DID Specification Registries and allows DID method authors to claim their entry. Presumably, this would solve the problem of keeping the directory up to date by relying on the method authors. However, at the time of writing, only a small minority of methods has been claimed.

\section{Methodology}
\label{sec:methodology}

We now elaborate on the methodology used to develop our proposed taxonomy. As we are developing a taxonomy in the information systems space, we chose to follow the approach publicized by Nickerson et al.~\cite{nickerson2013method}. They present an iterative approach towards developing a taxonomy that combines empirical and conceptual strategies. At this point, it is tried and tested and has rightfully established itself as the de facto default for taxonomy development in the information systems space.


\subsection{Meta-Characteristic}

One important corner stone of taxonomy development following Nickerson et al. is to define what is called a "meta-characteristic". It exists to give purpose to a taxonomy under construction and ensures that the choice of characteristics is coherent. While this can be chosen later on during the development of a taxonomy, we had a set goal from the start. We aim to enable practitioners to efficiently select DID methods to support and actively use with different stakeholders for new or upgraded software systems. More specifically, the practitioners we have in mind are mainly system architects and researchers building pilot systems. With this goal in mind, we define our meta-characteristic as follows:

The capabilities and usage characteristics of the DID methods, such as setup requirements, feature support, and cost.

We expect users of our taxonomy to have a basic understanding of what a DID is, but we do recognize that most practitioners whose systems could benefit from embracing decentralized standards do not have a strong background in decentralized identity. We instead assume them to be application-savvy. They know what they need and generally do not have time or motivation to do a technology deep-dive.


\begin{table*}[!t]
\centering
\caption{Objective Ending Conditions}
\setlength{\extrarowheight}{4pt}
\begin{tabularx}{\textwidth}{p{0.35\textwidth}X}
\toprule
\textit{Objective Ending Condition} &
  \textit{Comment} \\ \midrule
All sufficiently mature specifications from the Universal Resolver and the W3C Registries List have been examined. &
  Both of these sources together provide a pool of ca. 160 DID method specifications before curating them further. Even taking into account a significant rate of abandoned and incomplete specifications, that is an extensive and exhaustive sample size. \\
No object was merged with a similar object or split into multiple objects in the last iteration. &
  Changes to the structure of the taxonomy might have a ripple effect, requiring further changes before it is conforming to the ending conditions. Splitting objects might create a new dimension with a new characteristic for each resulting object. Then that new dimension needs to be further examined for additional characteristics.1 \\
At least one object is classified under every characteristic of every dimension. &
  Including characteristics that are not found in existing DID methods might be interesting from a research perspective to uncover gaps in the design space. However, we design a tool for practitioners that can only choose from what exists and is usable. \\
No new dimensions or characteristics were added in the last iteration. &
  Again, changes to the structure of the taxonomy might require adjustments to other parts of the taxonomy. Adding a new and important dimension might increase the size of the dimension to a degree where removal of another dimension of lesser importance should be considered. \\
No dimensions or characteristics were merged or split in the last iteration. &
  Merges or splits in characteristics and dimensions might reflect a change in targeted granularity of the taxonomy. This change of course needs to be propagated through the entire taxonomy before one can stop. \\
Every dimension is unique and not repeated. &
  Dimensions that add no or not enough value due to repeated information just needlessly bloat the taxonomy. \\
Every characteristic is unique. &
  Here we slightly deviate from Nickerson et al. by making the condition more strict. Each dimension should be distinct from the other dimensions, and thus there should never be overlap between the characteristics. That ensures an objective degree of conciseness. \\
Each combination of characteristics is unique to one object and not repeated. &
  This is an important condition as we want the taxonomy to be as manageable in size as possible. \\
\bottomrule
\end{tabularx}
\label{tab1}
\end{table*}

\begin{table*}[!t]
\centering
\caption{Subjective Ending Conditions}
\setlength{\extrarowheight}{4pt}
\begin{tabularx}{\textwidth}{lX}
\toprule
\emph{Subjective Ending Condition} &
  \emph{Comment} \\ \midrule
Concise &
  It should be possible to get some overview of the taxonomy upon first glance. We deliberated turning this into an objective ending condition by committing to a specific maximum of dimensions, but came to the conclusion that it might not be in the best interest of developing a useful taxonomy to arbitrarily limit the dimensionality. \\
Robust &
  The dimensions and characteristics should be chosen to provide meaningful separation between objects. Specifically, a practitioner must be able to gain actionable information from the classification of an object. \\
Comprehensive &
  We already define an objective measure of comprehensiveness regarding object inclusion as our first objective ending condition. Thus, subjective comprehensiveness is focused on the inclusion of dimensions. Specifically, we consider our taxonomy to be comprehensive if the dimensions cover key questions that a practitioner would have about a DID method. \\
Extendible &
  Keeping the rapid pace of development in the decentralized identity field in mind, we aim to construct our taxonomy in a way that makes it possible to add new dimensions or characteristics. Specifically, that entails the avoidance of unspecific bundling characteristics along the lines of  "all others". \\
Explanatory &
  The taxonomy's main purpose is to provide meaningful information about DID methods that indirectly explains the consequences of and requirements for using a given DID method. \\
Timeless &
  We want this taxonomy to be useful many years in the future and thus cannot focus on aspects that would be obsolete within months, given the pace of development. For example, that means we cannot include information regarding software support. \\
\bottomrule
\end{tabularx}
\label{tab2}
\end{table*}

\subsection{Ending Conditions}

The other corner stone of the methodology following Nickerson et al. is defining ending conditions. These have to be set at the very start to be referred back to in every iteration, until they are finally met by the completed taxonomy. They can be divided into objective and subjective ending conditions. The objective ones provide a clear picture of the process and resulting taxonomy. In contrast, the subjective ending conditions focus on the result only. We closely adapt our ending conditions from Nickerson et al. because they form a solid basis independent of meta-characteristic or specific fields. We slightly modify some of them and add one additional subjective ending condition. The resulting objective ending conditions are explained in Table \ref{tab1} and the subjective ones in Table \ref{tab2}.

\subsection{Defining Objects} \label{secObjects}

Up until now, we have cautiously avoided clarifying what exactly an \emph{object} is within the scope of this taxonomy. The first solution coming to mind is making each DID method an object. Or, later on, after grouping them, groups of DID methods. When setting out to create this taxonomy, we first spent some time examining any DID method specifications we could find. After going through roughly 160 methods and writing down some basic data such as the underlying registry and also observations we made along the way. It soon became obvious that it would not be feasible to meaningfully classify DID methods because some of them have an incredible range of possible DIDs within their specification. For example, there are method specifications defining several sub-methods or supporting vastly different registries. Especially the latter is very common and also appears in less obvious ways. The method \emph{did:ethr} is one of the obvious examples. The method was specifically engineered to support any Ethereum Virtual Machine-compatible blockchain as storage for identifier information that can subsequently be used to construct a DID document for a given DID, which contains a network identifier as part of the method specific identifier. That means the method covers networks ranging from the Ethereum mainnet, a public permissionless blockchain with significant transaction fees governed by Proof of Stake, to a small private networks using Proof of Authority and consequently having no transaction fees. These two extremes, and anything in between, have vastly different implications for a practical user.

A less obvious example of registry variability is the DID method \emph{did:web}. It allows a user to create a DID by hosting a DID document directly on a traditional web server addressed via a domain name or a specific URL. Usually, this entails buying a domain and paying for hosting. That creates an easy to understand, human-readable DID with some initial setup complexity and recurring cost. But, by using a public GIT service, such as GitHub, everyone can host a DID document for free with minimal effort, while giving up some control over the URL that becomes the DID. This, again, has vastly different implications for a potential user wanting to create such a DID. And trying to accurately reflect these different ways of using a DID method leads to dimensions that tend to explode into a large number of characteristics trying to capture\textemdash often contradicting\textemdash characteristic combinations as one.

For these reasons, we ultimately decided to approach the problem by subdividing DID methods. We refer to these subdivisions as \emph{DID method instances}. Each one refers to one specific usage pattern of a given DID method conforming, but not necessarily explicitly mentioned in, its specification.
The taxonomy we propose in this work is classifying objects that are groups of DID method instances. Every group has one associated DID method instance that is a prime example for the group. We never consider single DID method instances, even if we might only be aware of a single one within a group, because the DID methods, and thus DID method instances, might change a lot in the future. But the groups and their characteristics will most likely persist.



\subsection{Limitations}
Next, we want to address some limitations that our methodology imposes on our resulting taxonomy. First, we cannot include DID methods that we do not have sufficient information on. That means that all incomplete or imprecise DID method specifications were excluded from the taxonomy building process. Especially, company created DID methods are often lacking transparency about the underlying infrastructure. Also, some method specifications were not available in English and were consequently excluded as well. This should be a minor limitation, though, because a DID method with inadequate specification probably has no viable software support anyway. In addition, we have also excluded the few existing joke methods, such as \emph{did:did}. They were never meant or suited to be taken seriously by a practitioner.


Second, the commitment to creating a taxonomy that can serve as a long-lasting tool in a fast-moving space imposes some restrictions on feasible characteristics and thus dimensions. For this reason, we cannot include maturity levels or details on implementation support, as they evolve quickly over time.

Third, we may miss some possible DID method instances that we are not aware of. If they are not explicitly included in a DID method specification, it takes prior experience with the method, its underlying registry technology, and some creativity to infer all possible DID method instances. There is nothing we can do to specifically mitigate this. It is always a risk to miss objects when constructing a taxonomy. As we build with extensibility in mind, we are confident that small amounts of new DID method instances could easily be added in the future.

\subsection{Building the Taxonomy}
In total, we were able to identify around 160 DID methods. Many of them are in a very early phase of development, abandoned, or both. To start a meaningful taxonomy, we thus looked at the list of methods included in the universal resolver first. Still, we had to remove a few methods due to incomplete documentation. From there on, after the first empirical-to-conceptual iteration, we focused on conceptual-to-empirical iterations to further round out the range of dimensions and characteristics. Towards the end, we then circled back to empirical-to-conceptual iterations to examine all further DID methods we could find from open lists or via snowballing from specifications and documentations.

\section{A Proposed Taxonomy for DID Method Selection}
\label{sec:taxonomy}

\begin{table*}[!t]
\centering
\caption{Proposed Taxonomy for DID Method Selection}
\label{tabTaxonomy}
\setlength{\tabcolsep}{5pt}
\begin{tabular}{|l"c|c"c|c|c|c"c|c|c"c|c"c|c|c|c"c|c"c|c|c|c|c|} 
\multicolumn{1}{l}{\vcell{}}                                                                                                             & \multicolumn{2}{l}{\vcell{\begin{sideways}\textbf{Use~Case}\end{sideways}}}                   & \multicolumn{4}{l}{\vcell{\begin{sideways}\begin{tabular}[b]{@{}l@{}}\textbf{Registry}\\\textbf{Technology}\end{tabular}\end{sideways}}}                                                        & \multicolumn{3}{l}{\vcell{\begin{sideways}\begin{tabular}[b]{@{}l@{}}\textbf{Deployment}\\\textbf{Required}\end{tabular}\end{sideways}}}                                         & \multicolumn{2}{l}{\vcell{\begin{sideways}\begin{tabular}[b]{@{}l@{}}\textbf{Operation}\\\textbf{Support}\end{tabular}\end{sideways}}} & \multicolumn{4}{l}{\vcell{\begin{sideways}\begin{tabular}[b]{@{}l@{}}\textbf{Explicit}\\\textbf{Cost}\end{tabular}\end{sideways}}}                                                                                    & \multicolumn{2}{l}{\vcell{\begin{sideways}\begin{tabular}[b]{@{}l@{}}\textbf{Identifier}\\\textbf{Format}\end{tabular}\end{sideways}}} & \multicolumn{5}{l|}{\vcell{\begin{sideways}\begin{tabular}[b]{@{}l@{}}\textbf{DID Document~~}\\\textbf{Capabilities}\end{tabular}\end{sideways}}}                                                                                                   \\[-\rowheight]
\cline{2-23}
\multicolumn{1}{l"}{\printcelltop}                                                                                                        & \multicolumn{2}{l"}{\printcellbottom}                                                          & \multicolumn{4}{l"}{\printcellbottom}                                                                                                                                                            & \multicolumn{3}{l"}{\printcellbottom}                                                                                                                                             & \multicolumn{2}{l"}{\printcellbottom}                                                                                                   & \multicolumn{4}{l"}{\printcellbottom}                                                                                                                                                                                  & \multicolumn{2}{l"}{\printcellbottom}                                                                                                   & \multicolumn{5}{l|}{\printcellbottom}                                                                                                                                                                                                             \\
\hline
\multicolumn{1}{c}{\vcell{\begin{tabular}[b]{@{}c@{}}\textbf{DID Instance Group Name}\\\textit{Example Method Instances}\end{tabular}}} & \vcell{\begin{sideways}General\end{sideways}} & \vcell{\begin{sideways}Specific\end{sideways}} & \vcell{\begin{sideways}Self-Contained\end{sideways}} & \vcell{\begin{sideways}DLT\end{sideways}} & \vcell{\begin{sideways}Web Service\end{sideways}} & \vcell{\begin{sideways}DHT\end{sideways}} & \vcell{\begin{sideways}No Deployment\end{sideways}} & \vcell{\begin{sideways}Independent Deployment\end{sideways}} & \vcell{\begin{sideways}Coordinated Deployment\end{sideways}} & \vcell{\begin{sideways}CR\end{sideways}} & \vcell{\begin{sideways}CRUD\end{sideways}}                                                   & \vcell{\begin{sideways}Free\end{sideways}} & \vcell{\begin{sideways}Write Fee\end{sideways}} & \vcell{\begin{sideways}Recurring Cost\end{sideways}} & \vcell{\begin{sideways}Write Fee + Recurring Cost~~\end{sideways}} & \vcell{\begin{sideways}Not Human-Readable\end{sideways}} & \vcell{\begin{sideways}Human-Readable\end{sideways}}                         & \vcell{\begin{sideways}Minimal\end{sideways}} & \vcell{\begin{sideways}Basic\end{sideways}} & \vcell{\begin{sideways}Keys\end{sideways}} & \vcell{\begin{sideways}Services\end{sideways}} & \vcell{\begin{sideways}Arbitrary Data\end{sideways}}  \\[-\rowheight]
\multicolumn{1}{|c"}{\printcellmiddle}                                                                                                    & \printcellbottom                              & \printcellbottom                               & \printcellbottom                                     & \printcellbottom                          & \printcellbottom                                  & \printcellbottom                          & \printcellbottom                                    & \printcellbottom                                             & \printcellbottom                                             & \printcellbottom                         & \printcellbottom                                                                             & \printcellbottom                           & \printcellbottom                                & \printcellbottom                                     & \printcellbottom                                                 & \printcellbottom                                         & \printcellbottom                                                             & \printcellbottom                              & \printcellbottom                            & \printcellbottom                           & \printcellbottom                               & \printcellbottom                                      \\ 
\thickhline
\begin{tabular}[c]{@{}l@{}}Expressive Lightweight DIDs\\\textit{peer}\end{tabular}                                                        & x                                             & ~                                              & x                                                    & ~                                         & ~                                                 & ~                                         & x                                                   & ~                                                            & ~                                                            & ~                                        & x                                                                                            & x                                          & ~                                               & ~                                                    & ~                                                                & x                                                        & ~                                                                            & ~                                             & ~                                           & ~                                          & x                                              & ~                                                     \\ 
\hline
\begin{tabular}[c]{@{}l@{}}Simple Lightweight DIDs\\\textit{key, pkh}\end{tabular}                                                        & x                                             & ~                                              & x                                                    & ~                                         & ~                                                 & ~                                         & x                                                   & ~                                                            & ~                                                            & x                                        & ~                                                                                            & x                                          & ~                                               & ~                                                    & ~                                                                & x                                                        & ~                                                                            & ~                                             & x                                           & ~                                          & ~                                              & ~                                                     \\ 
\hline
\begin{tabular}[c]{@{}l@{}}Basic Consortium DLT DIDs\\\textit{ev:\textless{}consortium mnid\textgreater{}}\end{tabular}                   & x                                             & ~                                              & ~                                                    & x                                         & ~                                                 & ~                                         & ~                                                   & ~                                                            & x                                                            & ~                                        & x                                                                                            & x                                          & ~                                               & ~                                                    & ~                                                                & x                                                        & ~                                                                            & ~                                             & ~                                           & x                                          & ~                                              & ~                                                     \\ 
\hline
\begin{tabular}[c]{@{}l@{}}Basic Public DLT DIDs\\\textit{ev:\textless{}mainnet mnid\textgreater{}}\end{tabular}                          & x                                             & ~                                              & ~                                                    & x                                         & ~                                                 & ~                                         & x                                                   & ~                                                            & ~                                                            & ~                                        & x                                                                                            & ~                                          & x                                               & ~                                                    & ~                                                                & x                                                        & ~                                                                            & ~                                             & ~                                           & x                                          & ~                                              & ~                                                     \\ 
\hline
\begin{tabular}[c]{@{}l@{}}Custom DNS-based DIDs\\\textit{web:\textless{}mydomain\textgreater{}}\end{tabular}                             & x                                             & ~                                              & ~                                                    & ~                                         & x                                                 & ~                                         & ~                                                   & x                                                            & ~                                                            & ~                                        & x                                                                                            & ~                                          & ~                                               & x                                                    & ~                                                                & ~                                                        & x                                                                            & ~                                             & ~                                           & ~                                          & ~                                              & x                                                     \\ 
\hline
\begin{tabular}[c]{@{}l@{}}Free DHT-based DIDs\\\textit{oyd + public log server}\end{tabular}                                             & x                                             & ~                                              & ~                                                    & ~                                         & ~                                                 & x                                         & x                                                   & ~                                                            & ~                                                            & ~                                        & x                                                                                            & x                                          & ~                                               & ~                                                    & ~                                                                & x                                                        & ~                                                                            & ~                                             & ~                                           & ~                                          & ~                                              & x                                                     \\ 
\hline
\begin{tabular}[c]{@{}l@{}}Free Service Capable Public DLT DIDs\\\textit{ion + 3rd party anchor}\end{tabular}                             & x                                             & ~                                              & ~                                                    & x                                         & ~                                                 & ~                                         & x                                                   & ~                                                            & ~                                                            & ~                                        & x                                                                                            & x                                          & ~                                               & ~                                                    & ~                                                                & x                                                        & ~                                                                            & ~                                             & ~                                           & ~                                          & x                                              & ~                                                     \\ 
\hline
\begin{tabular}[c]{@{}l@{}}Free Human-Readable DIDs\\\textit{web:github.com:\textless{}path\textgreater{}}\end{tabular}                   & x                                             & ~                                              & ~                                                    & ~                                         & x                                                 & ~                                         & x                                                   & ~                                                            & ~                                                            & ~                                        & x                                                                                            & x                                          & ~                                               & ~                                                    & ~                                                                & ~                                                        & x                                                                            & ~                                             & ~                                           & ~                                          & ~                                              & x                                                     \\ 
\hline
\begin{tabular}[c]{@{}l@{}}Fully Capable Consortium DLT DIDs\\\textit{indy:\textless{}myconsortium\textgreater{}}\end{tabular}            & x                                             & ~                                              & ~                                                    & x                                         & ~                                                 & ~                                         & ~                                                   & ~                                                            & x                                                            & ~                                        & x                                                                                            & x                                          & ~                                               & ~                                                    & ~                                                                & x                                                        & ~                                                                            & ~                                             & ~                                           & ~                                          & ~                                              & x                                                     \\ 
\hline
\begin{tabular}[c]{@{}l@{}}Fully Capable DHT-based DIDs\\\textit{onion, gns, ipid}\end{tabular}                                           & x                                             & ~                                              & ~                                                    & ~                                         & ~                                                 & x                                         & ~                                                   & x                                                            & ~                                                            & ~                                        & x                                                                                            & x                                          & ~                                               & ~                                                    & ~                                                                & x                                                        & ~                                                                            & ~                                             & ~                                           & ~                                          & ~                                              & x                                                     \\ 
\hline
\begin{tabular}[c]{@{}l@{}}Free Fully Capable Layer 2 DIDs\\\textit{3 + public CAS}\end{tabular}                                          & x                                             & ~                                              & ~                                                    & x                                         & ~                                                 & ~                                         & ~                                                   & x                                                            & ~                                                            & ~                                        & x                                                                                            & x                                          & ~                                               & ~                                                    & ~                                                                & x                                                        & ~                                                                            & ~                                             & ~                                           & ~                                          & ~                                              & x                                                     \\ 
\hline
\begin{tabular}[c]{@{}l@{}}Fully Capable Permissioned DLT DIDs\\\textit{indy:sovrin, sov}\end{tabular}                                    & x                                             & ~                                              & ~                                                    & x                                         & ~                                                 & ~                                         & x                                                   & ~                                                            & ~                                                            & ~                                        & x                                                                                            & ~                                          & ~                                               & ~                                                    & x                                                                & x                                                        & ~                                                                            & ~                                             & ~                                           & ~                                          & ~                                              & x                                                     \\ 
\hline
\begin{tabular}[c]{@{}l@{}}Fully Capable Public DLT DIDs\\\textit{polygon, iota:main}\end{tabular}                                        & x                                             & ~                                              & ~                                                    & x                                         & ~                                                 & ~                                         & x                                                   & ~                                                            & ~                                                            & ~                                        & x                                                                                            & ~                                          & x                                               & ~                                                    & ~                                                                & x                                                        & ~                                                                            & ~                                             & ~                                           & ~                                          & ~                                              & x                                                     \\ 
\hline
\begin{tabular}[c]{@{}l@{}}Fully Capable Rented Public DLT DIDs\\\textit{sol}\end{tabular}                                                & x                                             & ~                                              & ~                                                    & x                                         & ~                                                 & ~                                         & x                                                   & ~                                                            & ~                                                            & ~                                        & x                                                                                            & ~                                          & ~                                               & ~                                                    & x                                                                & x                                                        & ~                                                                            & ~                                             & ~                                           & ~                                          & ~                                              & x                                                     \\ 
\hline
\begin{tabular}[c]{@{}l@{}}Human-Readable DLT NS DIDs\\\textit{ens:mainnet}\end{tabular}                                                  & x                                             & ~                                              & ~                                                    & x                                         & ~                                                 & ~                                         & x                                                   & ~                                                            & ~                                                            & ~                                        & x                                                                                            & ~                                          & ~                                               & ~                                                    & x                                                                & ~                                                        & x                                                                            & ~                                             & ~                                           & x                                          & ~                                              & ~                                                     \\ 
\hline
\begin{tabular}[c]{@{}l@{}}Human-Readable Name System DIDs\\\textit{dns}\end{tabular}                                                     & x                                             & ~                                              & ~                                                    & ~                                         & x                                                 & ~                                         & x                                                   & ~                                                            & ~                                                            & ~                                        & x                                                                                            & ~                                          & ~                                               & x                                                    & ~                                                                & ~                                                        & x                                                                            & ~                                             & ~                                           & ~                                          & x                                              & ~                                                     \\ 
\hline
\begin{tabular}[c]{@{}l@{}}Service Capable Consortium DLT DIDs\\\textit{ethr:\textless{}myconsortium\textgreater{}}\end{tabular}          & x                                             & ~                                              & ~                                                    & x                                         & ~                                                 & ~                                         & ~                                                   & ~                                                            & x                                                            & ~                                        & x                                                                                            & x                                          & ~                                               & ~                                                    & ~                                                                & x                                                        & ~                                                                            & ~                                             & ~                                           & ~                                          & x                                              & ~                                                     \\ 
\hline
\begin{tabular}[c]{@{}l@{}}Service Capable DHT-based DIDs\\\textit{orb + own log servers}\end{tabular}                                    & x                                             & ~                                              & ~                                                    & ~                                         & ~                                                 & x                                         & ~                                                   & ~                                                            & x                                                            & ~                                        & x                                                                                            & x                                          & ~                                               & ~                                                    & ~                                                                & x                                                        & ~                                                                            & ~                                             & ~                                           & ~                                          & x                                              & ~                                                     \\ 
\hline
\begin{tabular}[c]{@{}l@{}}Service Capable Public DLT DIDs\\\textit{ethr:mainnet}\end{tabular}                                            & x                                             & ~                                              & ~                                                    & x                                         & ~                                                 & ~                                         & x                                                   & ~                                                            & ~                                                            & ~                                        & x                                                                                            & ~                                          & x                                               & ~                                                    & ~                                                                & x                                                        & ~                                                                            & ~                                             & ~                                           & ~                                          & x                                              & ~                                                     \\ 
\hline
\begin{tabular}[c]{@{}l@{}}Generative Asset Identifying DIDs\\\textit{asset}\end{tabular}                                                 & ~                                             & x                                              & x                                                    & ~                                         & ~                                                 & ~                                         & x                                                   & ~                                                            & ~                                                            & x                                        & ~                                                                                            & x                                          & ~                                               & ~                                                    & ~                                                                & x                                                        & ~                                                                            & x                                             & ~                                           & ~                                          & ~                                              & ~                                                     \\ 
\hline
\begin{tabular}[c]{@{}l@{}}Content Code Claimant DIDs\\\textit{iscc}\end{tabular}                                                         & ~                                             & x                                              & ~                                                    & x                                         & ~                                                 & ~                                         & x                                                   & ~                                                            & ~                                                            & ~                                        & x                                                                                            & ~                                          & x                                               & ~                                                    & ~                                                                & x                                                        & ~                                                                            & ~                                             & ~                                           & ~                                          & ~                                              & x                                                     \\ 
\hline
\begin{tabular}[c]{@{}l@{}}Private Document Identifying DIDs\\\textit{schema:\textless{}private ipfs\textgreater{}}\end{tabular}          & ~                                             & x                                              & ~                                                    & ~                                         & ~                                                 & x                                         & ~                                                   & ~                                                            & x                                                            & x                                        & ~                                                                                            & x                                          & ~                                               & ~                                                    & ~                                                                & x                                                        & ~                                                                            & x                                             & ~                                           & ~                                          & ~                                              & ~                                                     \\ 
\hline
\begin{tabular}[c]{@{}l@{}}Public Document Identifying DIDs\\\textit{schema:public-ipfs}\end{tabular}                                     & ~                                             & x                                              & ~                                                    & ~                                         & ~                                                 & x                                         & ~                                                   & x                                                            & ~                                                            & x                                        & ~                                                                                            & x                                          & ~                                               & ~                                                    & ~                                                                & x                                                        & ~                                                                            & x                                             & ~                                           & ~                                          & ~                                              & ~                                                     \\ 
\hline
\begin{tabular}[c]{@{}l@{}}Specialized Consortium DLT DIDs\\\textit{hpass}\end{tabular}                                                   & ~                                             & x                                              & ~                                                    & x                                         & ~                                                 & ~                                         & ~                                                   & ~                                                            & x                                                            & ~                                        & x                                                                                            & x                                          & ~                                               & ~                                                    & ~                                                                & x                                                        & ~                                                                            & ~                                             & ~                                           & x                                          & ~                                              & ~                                                     \\ 
\hline
\begin{tabular}[c]{@{}l@{}}Specialized Fully Capable DLT DIDs\\\textit{panacea}\end{tabular}                                              & ~                                             & x                                              & ~                                                    & x                                         & ~                                                 & ~                                         & x                                                   & ~                                                            & ~                                                            & ~                                        & x                                                                                            & ~                                          & x                                               & ~                                                    & ~                                                                & x                                                        & ~                                                                            & ~                                             & ~                                           & ~                                          & ~                                              & x                                                     \\
\hline
\end{tabular}
\end{table*}

Our proposed taxonomy is shown in Table \ref{tabTaxonomy}. As elaborated on earlier in Section \ref{secObjects}, we classify groups of DID method instances. We have named these groups we formed from all of the different DID instances we examined, and tried to keep the names as short as possible while preserving descriptiveness. For illustrative purposes, we have also added one or more examples for every DID method instance group.
In the following sections, we first take a deeper look at our chosen dimensions with all of their characteristics. Then, we describe how a practitioner might use the taxonomy to select relevant DID methods for a project.

\subsection{Dimensions}

The taxonomy consists of seven dimensions with between two and five characteristics each. In the following, we go through all of them and discuss them in detail.

\begin{itemize}
\setlength{\itemindent}{-1em}
  \item[] \textbf{Use Case} This first dimension separates the DID method instances by intended use case. Because we expect most practical uses being served by general-purpose DIDs and because we want to preserve the approachability of the taxonomy, we have chosen to not distinguish between different specific use cases. Thus, the dimension has only these characteristics:
  \begin{itemize}
    \item General: designed for general-purpose use
    \item Specific: designed to serve a specific use case
  \end{itemize}
  \item[] \textbf{Registry Technology} The second dimension focuses on the technology of the registry used by a DID method instance. The registry is not necessarily what the DID document is written to, but rather whatever is holding the authority to establish consensus on what DID document is correct and, if the method supports, also at what time. That means that if multiple storages are used, such as \texttt{did:ion} using Ethereum to anchor and timestamp its DID document operations executed on the Ceramic network, we consider Ethereum to be the registry. We identified the following different types of registry:
  \begin{itemize}
    \item Self-Contained: no registry is needed as all information is derived from the DID itself
    \item DLT: the DID is registered on a distributed ledger
    \item Web Service: the DID is registered using some type of web service
    \item DHT: the DID is registered in a distributed hash table, such as the interplanetary file system (IPFS)
  \end{itemize}
  \item[] \textbf{Deployment Requirement} The next dimension describes the nature of potentially needed deployment to support the use of a certain DID method instance. It is important to note that we specifically address deployment, not hosting. If a deployment is needed, we make no statement on whether it is hosted directly by a practitioner or contracted out in some way. The dimension only states whether a personal deployment of some sort is required to adopt a DID method instance. The characteristics are as follows:
  \begin{itemize}
    \item No Deployment: some registries, such as public blockchains, can be accessed via existing deployed nodes and do not require a new deployment
    \item Independent Deployment: a personal deployment that can be independently set up, such as an interplanetary file system (IPFS) node, is required
    \item Coordinated Deployment: a personal deployment is required and needs to be coordinated with other new deployments by cooperating parties, as is the case for a consortium blockchain
  \end{itemize}
  \item[] \textbf{Operation Support} DID methods can support different ranges of the classic CRUD operations: Create, Read (or Resolve), Update, Delete (or Deactivate). Only the first two operations are mandatory due to them being required for any meaningful use of a DID. Delete is a special case when it comes to DIDs. Depending on the registry technology, it might not be possible to delete any data. For example, a public blockchain with its permanent record of history would prevent real data deletion. Thus, we consider the last CRUD operation to be Deactivate. For the majority of use cases, it should be sufficient to indicate deactivation of an identity. Personally identifiable information that would call for real deletion should never be written on a public blockchain in the first place. We observed two different sets of supported operations leading to these characteristics:
  \begin{itemize}
    \item CR: creating and resolving are supported
    \item CRUD: the full spectrum of CRUD operations is supported
  \end{itemize}
  \item[] \textbf{Explicit Cost} Apart from any cost incurred through deployment, which we have addressed in a previous dimension, there may be fixed costs associated with the adoption of a DID method instance, such as transaction fees on a public distributed ledger. It should be noted that these costs do not necessarily have to be paid by the DID controller. Depending on the exact use case, it might be desirable that a third party covers these. Because these fees can vary heavily from registry to registry and also over time, we do not categorize the cost by concrete amounts, but rather by cost model. For that, we have defined the following characteristics:
  \begin{itemize}
    \item Free: no cost is incurred
    \item Write Fee: fees have to be paid for write operations, which usually include creating, updating, and deactivating a DID
    \item Recurring Cost: fees have to be paid continuously for the usage duration of a DID, for example in the form of a subscription
    \item Write Fee + Recurring Cost: fees for write operations have to be paid on top of some continuous cost
  \end{itemize}
  \item[] \textbf{Identifier Format} The DID method specific part of a DID can take many different forms leading to different length and complexity of identifiers. In some cases, it might be useful or even necessary for humans to interact with the identifier. Therefore, we introduce the following two characteristics:
  \begin{itemize}
    \item Not Human-Readable: the identifier is based on a format not intuitive for humans, such as some type of hash of a public key
    \item Human-Readable: the identifier can be read, spoken, and remembered with relative ease
  \end{itemize}
  \item[] \textbf{DID Document Capabilities} The W3C DID standard itself imposes no limits on how much or what kind of data a DID document may contain. However, depending on design intent and registry infrastructure, the kinds of information that can be managed through a specific DID method instance are often limited. We have structured the characteristics representing this as follows:
  \begin{itemize}
    \item Minimal: the DID document is purely descriptive and does not contain any key material
    \item Basic: the DID document contains exactly one key that cannot be rotated
    \item Keys: the method supports full key management
    \item Services: the method supports full key management and service descriptions
    \item Arbitrary Data: the method supports managing arbitrary data as part of the DID document
  \end{itemize}
\end{itemize}

\subsection{Using the Taxonomy}

While this taxonomy might be useful to different kinds of users, we primarily envision practitioners to benefit from using it. Software architects working on a system that needs DIDs will likely encounter two general design questions regarding DID method support. First, they will have to choose one or two DID methods to be used by a select few actors critical to the use case. An example for this might be universities that are supposed to use their DIDs to issue digital diplomas in the form of W3C Verifiable Credentials (VCs). The other question pertains to what DID methods to support for the new system. Until a true universal resolver exists, any system can only support a limited number of DIDs, which should be chosen with the needs of the general user base taken into account.

A practitioner having limited experience with DIDs, which is likely given the relatively young age of the standard, will be faced with a steep learning curve. Consulting this taxonomy should help to quickly understand the possible differences between DID methods from a user perspective. Armed with a basic understanding of DIDs and project requirements, a practitioner can follow these steps:

\begin{enumerate}
    \item Using the project requirements, the practitioner can iteratively exclude DID method instances with certain characteristics.
    \item Having narrowed down the candidate pool of DID method instances to just a few, the practitioner can look into the provided examples to better understand the possible choices for their specific use case.
    \item Having an understanding of what methods are suitable for their project, the practitioner is equipped to branch out on their own and find further relevant methods. For these, the practitioner can also look into point-in-time dependent characteristics like maturity of the method and software support, before ultimately making an informed decision.
\end{enumerate}


\section{Conclusion}
\label{sec:conclusion}
This paper proposes a taxonomy of DID method instances designed to support practitioners in navigating the large space of possible DID methods to choose from. In comparison to previous works, we have taken a look at the full DID method landscape, explicitly adopted the viewpoint of a practitioner, and taken care to avoid time dependent characteristics for the sake of this work's longevity. Consequently, this taxonomy is designed to be the first and most important\textemdash rather than the only\textemdash tool a practitioner can use to then dive into focused review of timely material, such as documentation. We also anticipate our work to be useful to a wider audience, including but not limited to DID method creators, regulators, and web3 enthusiasts. The field of Decentralized Identifiers is still young, and comprehensive high-level overviews of this complex DID method space have not existed before.

We see several opportunities for further research. First, we have purposefully been as value neutral as possible about the DID methods instances and their characteristics in this work. But, with DIDs being an integral component of Self-Sovereign Identity (SSI) and there being a general push for decentralization of the web, it might be of interest to explicitly evaluate characteristics, such as grade of decentralization, registry infrastructure longevity, and level of tamper resistance.

Next, one could expand this work into a fully developed DID method selection methodology based on a decision tree. Because the dimensions of this taxonomy are chosen to answer important questions for adoption, the taxonomy could function as the basis for an interactive question catalog complemented by an index of up-to-date DID methods by instance group. The final artifact could be a standalone tool that enables a practitioner to find suitable DID methods for their use case.

During this work, we experienced first-hand how little standardization there still is for DID method specifications. That includes vast differences in format, completeness of information, information density, and even versioning of the specification. Any work contributing to making DID method specifications themselves more approachable would be valuable.

Finally, the taxonomy might benefit from being amended at a later point in time when the DID landscape has evolved further. New methods might require new groups and characteristics to be adequately classified.

\section*{Acknowledgment}
This work has been funded by the German Federal Ministry of Education and Research (BMBF) under grant M534800. The responsibility for the content of this publication lies with the authors.

\bibliographystyle{IEEEtran}
\bibliography{bibliography}

\end{document}